\documentstyle{laa}     

\begin{document}

   \thesaurus{ 11.07.1;  
               11.17.3;  
               12.07.1;  
             }
   \title{A further discussion on quasar-galaxy associations
          from gravitational lensing}


   \author{Xiang-Ping Wu$^{1,2}$}

   \offprints{Xiang-Ping Wu$^1$}

   \institute{$^1$DAEC, Observatoire de Paris-Meudon, 92195 Meudon Principal
Cedex,
               France\\
              $^2$Beijing Astronomical Observatory, Chinese Academy of
Sciences,
                Beijing 100080, China}

   \date{Received August 12, 1993; accepted \qquad\qquad 1993}

   \maketitle

   \begin{abstract}
%
Quasar-galaxy associations, if they result from the effect of gravitational
lensing by foreground galaxies, depend sensitively on the shape of the
quasar number counts. Two kinds of quasar number-magnitude relations are
predicted to produce quite different properties in quasar-galaxy associations:
the counts of Boyle, Shanks and Peterson (1988; BSP) provide both positive
and ``negative" associations between distant quasars and  foreground galaxies,
relating closely with the knee ($B\approx19.15$)
in these counts. However, Hawkins and
V\'eron (1993; HV) quasar data lead to only a positive magnitude-independent
quasar-galaxy association.  The current
observational evidence on quasar-galaxy associations, either positive
or null, is shown to be the natural result of gravitational lensing
if quasars follow the BSP number-magnitude relation. On the other hand,
the HV counts are unable to produce the reported
associations by the mechanism of gravitational lensing.
It is emphasized that  special attention should be paid to the limiting
magnitudes in the selected
quasar samples when one works on  quasar-galaxy associations.
   \end{abstract}

%

\section{Introduction}

One of the important consequences of gravitational lensing,
as first realized by  Gott \& Gunn (1974) before the discovery of
the first lensed quasar pair,
is that the surface number density of quasars near foreground galaxies
would be increased (denoted by the quasar enhancement $q_Q$)
because the distant quasars lying behind galaxies
would be magnified by lensing effect of the galaxies and then enter into
the detection limit (see also Canizares, 1981; Vietri \& Ostriker, 1983;
Schneider, 1986, 1987; Kovner, 1989; etc.). Equivalently, an
overdensity of foreground galaxies around high redshift quasars would
also exist (described by the galaxy enhancement $q_G$)(Schneider, 1989).
The first statistical evidence on such quasar-galaxy
associations was reported by Webster et al. (1988). They claimed a
significant enhancement of quasar surface density in the vicinity of galaxies.
Moreover, there have been many observations of an increased
number density of foreground galaxies towards distant quasar positions
[for a review, see Narayan (1992)] and most of the authors  confirm
the galaxy number excess around background quasars.
If such associations are real, gravitational lensing indeed provides
a natural explanation.

However, the present observational status on the associations seems to be far
from satisfactory.  Table 1 summarizes the results reported in the
two recent lensing meetings (Hamburg 1991 and Li\`ege 1993). It is noticed
that (1) $q_G$ found
by Magain and Van Drom has been decreased relative to the value quoted by
Narayan (1992) and (2)two authors (Kedziora-Chudczer and Yee) find no
evidence for foreground galaxy enhancement.
Although some suggestions have been made to improve
the confidence of the different results by choosing the same objects,
cross-calibrating the different observing techniques and using the same
criteria,   large samples and considerable  observing time are required
to further confirm the existence of quasar-galaxy associations.

   \begin{table}
      \caption{Foreground galaxy enhancement $q_G$}
         \label{ }
      \[
         \begin{array}{c|c|c|c|c}
            \hline
            {\rm authors}  & {\rm QSO \;No.} & {\rm selections}
			   & \theta\; {\rm range} (") & q_G\\
            \hline
	   {\rm Crampton} & 101 & V<18.5 & 0-6 & 1.4\pm0.5\\
	                  &     & z>1.5  &     &          \\
	    \hline
	   {\rm Kedziora-} & 181 & V<18.5 & 6-90 & \sim 1\\
	   {\rm Chudczer} &     & z>0.65 &    &      \\
	    \hline
           {\rm Magain} & 153 & \overline{V}=17.4 & 0-3 & \sim2.8\\
                        &     & \langle z\rangle=2.3 &   &   \\
	    \hline
	   {\rm Van\; Drom} & 136 & \overline{V}=17.4 & 3-13.7 & \sim1.46 \\
                          &     & \langle z\rangle=2.3 &     &      \\
	    \hline
	   {\rm Webster} & 68 & V<18 & 3-10  & \sim2\\
                         &    & 0.7<z<2.3 &   &  \\
	    \hline
	   {\rm Yee} &    94  & V<19 & 2-6 & 1.0\pm0.3 \\
               &        &     z>1.5       & 2-10 & 1.0\pm0.2\\
               &        &                 & 2-15 & 0.9\pm0.1\\
	    \hline
         \end{array}
      \]
   \end{table}
Nevertheless, observations have already provided  some important
implications for quasar-galaxy associations, and some valuable information
can be obtained if one reaches a better understanding of their mechanism --
gravitational lensing. The second effect of gravitational lensing is the
area distortion which reduces the magnitude of the first effect
(magnification),
leading to a lower surface number density of sources
(quasars) (Gott \& Gunn, 1974; Peacock, 1986; Narayan, 1989).
Consequently, a large enhancement is hard to reach.  All
the previous authors have paid their attentions to how to find the maximum
enhancement factor $q_Q$ or $q_G$.
In 1989 Narayan (1989) developed an elegant
formula for the calculation of $q_Q$ and Kovner (1989) discussed the
upper bound on the evaluation of $q_Q$.
However, one important fact has been neglected: that
null and even ``negative" associations
between high redshift quasars and
low redshift galaxies would also occur in some cases. That is to say,
instead observing an overdensity of quasars around foreground galaxies, one
may detect a number decrease of quasars in the vicinity of
foreground galaxies, i.e., $q_Q<1$.
In fact, Narayan (1989) has showed such an example
in his figure with $B_0=20$ and Kovner (1990) has also mentioned the case of
$q_Q<1$ for  fainter counts,
but they both didn't further discuss this effect.
As will be shown below, these null and/or ``negative" associations
play  an important role in the understanding of the observed associations
between quasars and galaxies.

%

\section{The dependence of $q_Q$ on quasar number counts}

The quasar enhancement factor $q_Q$ is the ratio of the observed quasar
surface number density around foreground galaxies
to their background mean value, which depends
on  (1)quasar number counts $N(<m)$ and
(2)magnification  $A$ by lensing effect of the galaxies (Narayan, 1989)
\begin{equation}
q_Q=\frac{N(<m+2.5\log A)}{N(<m)}\frac{1}{A},
\end{equation}
where $2.5\log A$ is the increased apparent magnitude by
the first effect of lensing
which leads to picking up the fainter quasars around foreground galaxies,
and $1/A$ represents
the factor of the distorted area by second lensing effect of the
galaxies, resulting in a decrease of quasar number density.

Since $q_Q$ is closely related to the quasar number-magnitude relation $N(<m)$,
a precise determination  of $N(<m)$  is required
for the theoretical study of quasar-galaxy associations.
On the other hand, one may test the reported $N(<m)$ in
different observations by using the evidence of quasar-galaxy associations.
Essentially, two kinds of quasar number-magnitude relation have been thus
far suggested from, respectively,
the survey of Boyle, Shanks and Peterson (1988) (BSP) and the
survey of Hawkins and V\'eron (1993) (HV).
The significant difference in these two relations is that the slope
of the BSP counts changes at $B\approx19.15$ from $0.86$ to $0.28$
while there is no such a turnover in the range of
$B\leq21$ in the HV counts.   Interestingly, the subsequent
surveys by the BSP group, including the recent X-ray observations,
have all confirmed the existence of the
flatting of $N(<m)$ at fainter magnitude (Boyle et al., 1990;
Boyle, Jones \& Shanks, 1991; Zitelli et al., 1992;  Boyle et al., 1993).

We will adopt these two kinds of quasar number-magnitude relations in the
following computations and compare their properties in the resulted
quasar-galaxy associations.
The BSP cumulative counts have been fitted by Narayan (1989) to be
\begin{equation}
\begin{array}{ll}
N(<B)=4.66\times10^{0.86(B-19.15)},  \;\;\;& B<19.15;\\
N(<B)=-10.95+15.61\times 10^{0.28(B-19.15)}, \;\;& B>19.15.
\end{array}
\end{equation}
This relation is valid for $z\leq2.2$ and $B<21$.
However, a later observation (Boyle, Jones \& Shanks, 1991) indicates that
eq.(2) holds true also to $B\leq22$.
The HV cumulative counts can be fitted by
\begin{equation}
N(<B)=6.25\times10^{0.51(B-19.15)}.
\end{equation}
This relation has been written in such a form that its slope
can be directly compared with the corresponding values of the BSP counts.

The enhancement $q_Q$ is plotted against the magnification $A$  in Fig.1
for both BSP and HV counts.  One should pay a special attention to
the case where $m+2.5\log A$ is larger than the quasar sample threshold.
For instance, the BSP and the HV surveys were
limited to $B\leq21$. Therefore, the number-magnitude relation
$N(<B+2.5\log A)$ would fail when $B+2.5\log A>21$, and
one cannot calculate the enhancement $q_Q$ beyond the survey limit:  this
sets a strong constraint
on the application range of eq.(1).  Thus, the conclusions of Narayan (1989)
should be taken cautiously, and in strict sense
his arguments are valid
only for these solid lines in Fig.1 rather than the whole magnification range.
The extrapolation of the solid lines requires the knowledge of  fainter
quasar counts.

   \begin{figure}
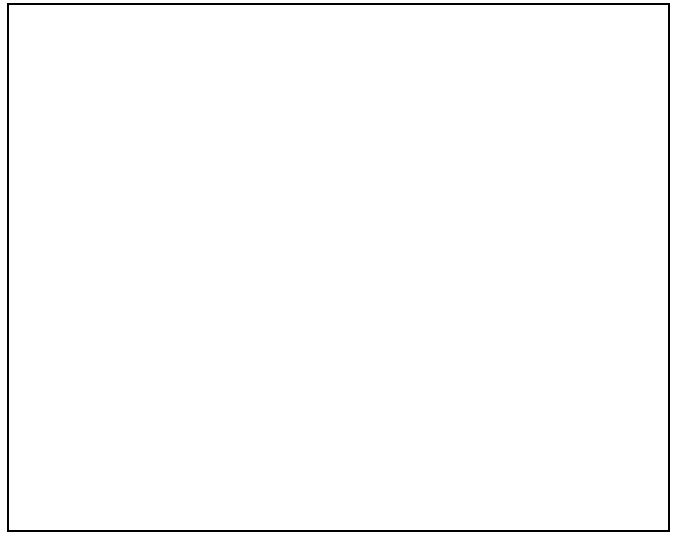

      \picplace{7cm}
      \caption{
The quasar enhancement $q_Q$ against the quasar limiting magnitude
$B$ and the
lensing magnification $A$ for the BSP and the HV counts. The solid lines
correspond to the results within the survey threshold of $B=21$
and the dotted lines are the extrapolated results of $N(<B)$
beyond $B=21$. Note that in the HV counts, $q_Q\geq1$, providing always
the positive associations, while in the BSP
counts $q_Q$ may be smaller than 1 when $B>19$,
resulting in the ``negative" associations.}
         \label{Fig.1}
    \end{figure}

A power-law number-magnitude relation  with index of $\alpha$,
$N(<m)\sim 10^{\alpha m}$,
would lead to an enhancement of $A^{2.5\alpha-1}$, independent of
the limiting magnitude. The HV
counts show that $\log q_Q/\log A=0.3$.  Hence, it is unlikely that a large
enhancement factor $q_Q$ can be reached in the HV counts.
For example, $q_Q>4$ requires the lensing magnification to be greater
than 100, which cannot be produced by a normal galaxy.
Nevertheless, the HV counts do not result in
the ``negative" associations, i.e.,  $q_Q\geq1$.

An example of the ``negative" association with $q_Q<1$ has been shown in Fig.1,
the curve with $B=20$ from the BSP counts. In fact, in the
range of $1<A<100$ the negative cases may occur for the limiting
magnitude $B>19$. The reason for this is that the BSP counts turn to be
strongly flatter after $B=19.15$, which would depress significantly the effect
of ``picking up" the fainter sources, while the second effect of lensing
remains unchanged.
As a result, one may observe a ``negative"
association between the background quasars and the foreground galaxies.
Note that $q_Q$ is not very sensitive to the selection
of the brightness of the foreground galaxies which
contribute only lensing  magnifications.

%

\section{Galaxies as deflectors}

For simplicity we adopt a singular isothermal sphere for
matter distribution in a lensing galaxy. The magnification in this
model can be simply written as
\begin{equation}
A=\frac{\theta}{\theta-\theta_c},
\end{equation}
where $\theta$ is the distance to the center of the galaxy on the galaxy plane
and $\theta_c$, the Einstein radius
\begin{equation}
\theta_c=4\pi \left(\frac{\sigma}{c}\right)^2\frac{D_{ds}}{D_s}.
\end{equation}
Here $\sigma$ is  velocity dispersion in the galaxy. $D_{ds}$ and $D_s$
measure the angular diameter distances from the galaxy and from the
observer to the distant source, respectively. The observations
which search for
quasar-galaxy associations often choose  galaxies at relatively lower
redshift ($z<0.3$) and quasars at higher redshift ($\langle z\rangle
\approx2$).  Therefore, the distance parameter $D_{ds}/D_s$ is nearly
unity. This is particularly true for  the nearby galaxies.
Anyhow, considering the actual distances does not
significantly change the following calculations. For a typical nearby galaxy,
\begin{equation}
\theta_c=1".52 \left(\frac{\sigma}{230{\rm km/s}}\right)^2.
\end{equation}

Fig.2  shows the enhancement $q_Q$ against the search ranges around
foreground galaxies at different limiting  magnitudes $B$.
It has been presumed that the extrapolation of both BSP and HV counts
to the fainter magnitude ($B>21$) is reasonable.
   \begin{figure}
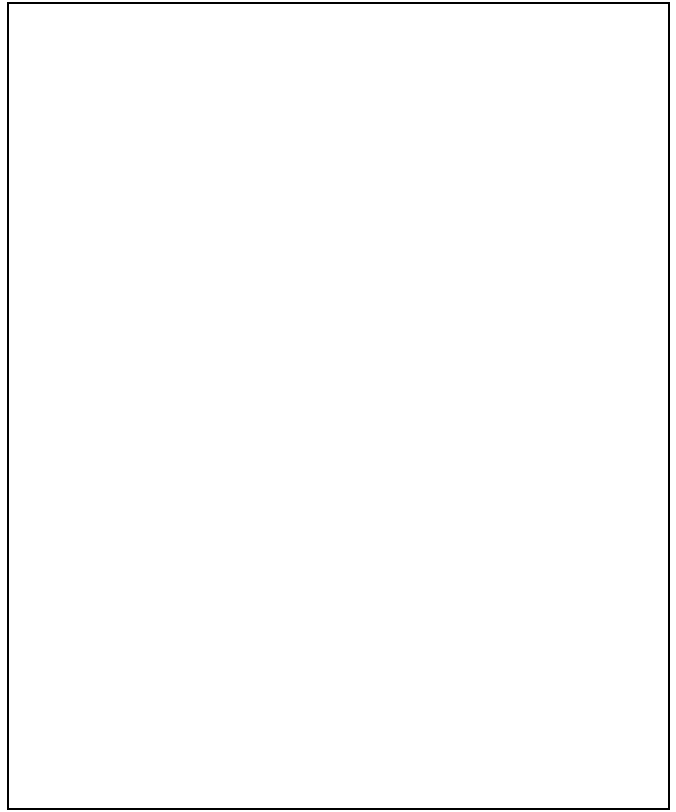

      \picplace{10.7cm}
      \caption{
The quasar enhancement factors $q_Q$ over different search areas around a
foreground galaxy with $\sigma=230$ km/s.
The curves have been extrapolated to the  fainter
magnitudes.}
         \label{Fig.2}
    \end{figure}
Since $q_Q$ is independent of limiting magnitude in the HV counts, the
resulted $q_Q$ relates only with the search areas. For a typical galaxy,
$q_Q$ is relatively small even when one focuses on the area very close to
the galaxy center.  Therefore, at least two results shown in Table 1 can
not be explained by the HV counts, Magain's $q_Q\sim2.8$
within $3"$ and Webster's
$q\sim2$ in the radius of $3-10"$. One may argue that a larger velocity
dispersion may provide a higher magnification, leading to the increase of
$q_Q$ in terms of eq.(1).  However, this would  simultaneously
produce a larger Einstein radius [eq.(6)], within which the
primary image of the lensed quasar does not appear. The occurrence of
the secondary image within the Einstein radius
is accompanied by the multiple images, which are very rare in
searches for the quasar-galaxy associations.

The most interesting result from the BSP counts is that both positive
and ``negative" associations can occur, separated in the range of
$19<B<20$.  Several important conclusions  are as follows: (1)The positive
associations between foreground galaxies and background quasars
would be found when one chooses the threshold of the quasar sample
to be brighter than $B\approx19.5$.
In particular, $q_Q$ is nearly independent of the
limiting magnitude $B$  if $B$ is smaller than 18.5 for $\theta<3"$ and
19 for the larger search areas around  galaxies.  (2)When the fainter
quasars ($B>19.5$) are involved in the sample for the purpose of finding the
quasar-galaxy associations, one would expect to detect null and/or
``negative" associations.
Similar to the positive cases, the strong ``negative" association ($q_Q<1$)
would be found if one looks for the ranges very close to the galaxy centers.
Note that the values of $q_Q$ in all the fainter cases are actually close to
1 and
therefore, one would simply have $q_Q\approx1$, i.e.,
the null association, if errors in the observations were significant.

Certainly, Fig.2 is only for a galaxy with a constant velocity dispersion
of $230$ km/s.
A more precise treatment is to consider galaxy luminosity distribution,
 e.g., the Schechter luminosity function,
and then to derive the galaxy distribution in velocity dispersion by
using a luminosity-velocity dispersion law, e.g.,
the Faber-Jackson relation.  We notice, however, that  different galaxies
can only lead
to the vertical shifts in amplitude in Fig.2 and the main features would
remain unchanged.

%

\section{Explanations of the present evidences}

It is generally believed that the foreground galaxy enhancement
factor $q_G$ should be in principle equal to the background
quasar enhancement factor $q_Q$.
The results on quasar-galaxy associations ($q_G$) listed
in Table 1, despite their large discrepancies, are in fact
consistent with the lensing predictions ($q_Q$) from the BSP quasar
number counts.  To see this, one
needs to translate the $V$ magnitudes in the observational studies of
quasar-galaxy associations into the $B$ magnitudes in the theoretical
predictions from lensing.  We take $\overline{(B-V)}\approx0.4$
for distant quasars, as suggested by P. V\'eron (private communication).

The Crampton  result of $q=1.4\pm0.5$ was obtained with a $B$ limiting
magnitude of $\sim18.9$. The BSP prediction in Fig.2 provides
the same value of $\sim1.4$ within $\theta=6"$ at $B\sim19$.

Kedziora-Chudczer kept the same limiting magnitude as Crampton but
searched
a larger range of separations. One can see from Fig.2 that around $B\sim19$
an average value over $\theta=6$--$90"$ tends to close to unity, roughly
consistent with the observation.

Magain's value is the highest one, $q=2.8$. However, he chose the brightest
quasars and the smallest search range ($\theta\leq3"$). At $B\sim18$ and
$\theta\leq3"$, lensing predicts that $q=2.3$. Considering the large
error bar in his data (Narayan, 1992), the reported $q$ is within the
prediction
of Fig.2.

With the same limiting magnitude, Van Drom's observations were made for
a larger radius. His finding of $q=1.46$ fits quite well to the value
in Fig.2.

Between $\theta=3"$ and $\theta=10"$, Webster found that $q\sim2$ at
$B_{limit}\sim18.4$. The predicted results are $1.3<q<2.3$
over these areas.

Finally Yee's negative or null evidences are the natural results of the faint
limiting magnitude used in his quasar samples ($B\sim19.4$).
As one can clearly see, around $B=19.5$ the ``turnover" occurs
despite  the search radii. Therefore,
the enhancements are expected to be roughly unity (i.e., the null or
``negative"
associations)  for all of his observations.

However, it should be mentioned that the threshold in each quasar
sample for searches of quasar-galaxy associations was not actually
very well defined and setting a clear limiting magnitude in each observation
is rather difficult.  Additionally, using a mean value for the limiting
magnitude in the survey often makes it hard to compare with the theoretical
predictions.  Moreover, the presently known  enhancement factors may contain
large errors and uncertainties, and therefore,
the above explanations of their gravitational
lensing origin still needs to be further investigated.

%

\section{Conclusions}

BSP and HV quasar counts show significantly different
behaviours in  quasar-galaxy associations, providing an efficient way to
test the proposed quasar number-magnitude relations.
Consequently, the present observational
evidences for these associations seem
to contradict the predictions from the HV counts while they are very well
fitted by the BSP data, confirming the overturn at $B\approx19.15$
in quasar counts. However, this conclusion should be considered
to be preliminary, depending on the significances of the reported
enhancement factors for quasar-galaxy associations.

The gravitational lensing origin of quasar-galaxy associations has been
found to be the natural explanations for all the reported evidences, either
positive or null (negative).  It is pointed out that the quasar limiting
magnitude in the survey for quasar-galaxy associations plays an important
role: The positive association would be found for the $B_{limit}<19.5$
quasar samples and the negative result,  for the $B_{limit}>19.5$
ones.  This conclusion is independent of the search ranges, no matter
how close the galaxies (quasars)  are chosen to the quasar (galaxy)
positions.

It is expected that the fainter quasar samples ($B>20$) would provide
some further evidences on the gravitational lensing origin of
quasar-galaxy associations. The confirmation
of the ``negative" associations between foreground galaxies and faint
background quasars would set very useful constraints on the shape of quasar
number-magnitude relation. Moreover, eq.(1) is actually a universal
relation for finding the enhancement by gravitational lensing. A future
work is to apply eq.(1) for background galaxies, which may provide very
useful information on 3CR galaxy-galaxy associations (Hammer \& Le f\`evre,
1990),  3CR galaxy-cluster associations (Roberts, O'Dell \& Burbidge,
1977; Hammer \& Le f\`evre, 1990)
and also the possible optical galaxy-cluster
associations, especially for K-selected galaxies.
Recall that the associations, either positive or negative,
between foreground objects and background sources may occur
if the slope of $\log N/\log m$ for background sources
is not exactly equal to $0.4$.

   \acknowledgements
I would like to thank Francois Hammer and Israel Kovner
for valuable comments and suggestions, and Michael Drinkwater and
Gary Mamon for careful reading the manuscript.
This work was supported by CNRS and Wong K.C. Foundation.

%

\end{document}